\documentclass[conference]{IEEEtran}
\IEEEoverridecommandlockouts
\usepackage{cite}
\usepackage{amsmath,amssymb,amsfonts}
\usepackage{algorithmic}
\usepackage{graphicx}
\usepackage{textcomp}
\usepackage{xcolor}
\usepackage[table]{xcolor}
\usepackage{placeins}
\usepackage{tabularx}
\usepackage{booktabs}
\usepackage{float}
\usepackage{array}
\usepackage{ragged2e}
\def\BibTeX{{\rm B\kern-.05em{\sc i\kern-.025em b}\kern-.08em
    T\kern-.1667em\lower.7ex\hbox{E}\kern-.125emX}}
\begin{document}

\title{Understanding NVIDIA Grace–Blackwell Unified Memory Through Placement-Controlled Microbenchmarks}

\author{
\IEEEauthorblockN{Nathan Graddon}
\IEEEauthorblockA{
\textit{University of Delaware}}
\and
\IEEEauthorblockN{Aaron Jarmusch}
\IEEEauthorblockA{
\textit{University of Delaware}}
\and
\IEEEauthorblockN{Sunita Chandrasekaran}
\IEEEauthorblockA{
\textit{University of Delaware}}
}

\maketitle

\begin{abstract}
NVIDIA Grace--Blackwell systems make CPU--GPU memory increasingly unified at the programming level, but unified addressability does not make the memory system uniform. A CUDA managed allocation on GB200 may be served from GPU-local HBM, accessed through host/system memory, demand-migrated, prefetched, reused after GPU access, or evicted and reloaded under capacity pressure. The problem is that these states share a programming interface while exercising different hardware paths and runtime mechanisms, making a single ``Unified Memory bandwidth'' number misleading.

This paper presents a placement-controlled microbenchmark study of CUDA Unified Memory on NVIDIA GB200, using GH200 as a Grace--Hopper baseline and GB10/DGX Spark as a validation platform. Our methodology separates local HBM access, host/system-memory access, explicit movement, managed first touch, prefetch, reuse, access-pattern sensitivity, oversubscription, and CPU--GPU handoff, then summarizes the observations with a measurement-calibrated model of compute, access, migration, eviction, synchronization, and handoff costs. On GB200, a 1~GB STREAM-style triad reaches 3963~GB/s from local HBM but 304~GB/s through host-mapped CPU/system memory; for a 256~MB streaming kernel, cold CPU-initialized managed memory reaches 14.1~GB/s, while managed prefetch and reuse reach 4606~GB/s and 4504~GB/s. Below the capacity boundary, prefetched and reused managed memory remain above 4.6~TB/s, but at 125\% of CUDA-visible memory they fall to 968~GB/s and 1214~GB/s, showing that Grace--Blackwell unified memory must be understood through placement, access history, locality, reuse, and capacity pressure.
\end{abstract}


\section{Introduction}
Modern heterogeneous systems increasingly expose memory abstractions that reduce the burden of explicit CPU--GPU data movement. CUDA Unified Memory (UM) allows host and device code to access one managed allocation while the runtime
handles mapping, migration, prefetching, and placement \cite{NVIDIA_CUDA_UM_2025,Harris2013CUDA6UM}. At the same time, tightly coupled platforms have changed the hardware on which these abstractions operate. NVIDIA Grace--Hopper introduced coherent CPU--GPU connectivity through NVLink-C2C, while Grace--Blackwell combines Grace CPU memory,
Blackwell GPU HBM, and larger NVLink-based configurations such as GB200 NVL72 \cite{NVIDIA_GraceHopper_Architecture,NVIDIA_NVLINK_C2C, NVIDIA_GB200_NVL72,NVIDIA_NVLINK_SYSTEMS}. These systems simplify CPU--GPU sharing, but they do not make all memory accesses equivalent.

The main issue is that unified addressability does not imply uniform performance. A managed allocation may correspond to data resident in GPU-local HBM, accessed through host/system memory, migrated on first touch, prefetched before execution, reused after GPU placement, or reloaded under oversubscription. These cases exercise different hardware paths and runtime mechanisms. Performance therefore depends on residency, initialization,
access pattern, reuse, and whether the working set exceeds GPU memory capacity.

Prior studies have shown that CUDA UM can reduce programming complexity while introducing costs from page faults, migration, prefetching, memory advice, and oversubscription \cite{Li2015UnifiedMemory,Chien2019CUDAUM,Sakharnykh2017UM, Allen2021UVM,Kim2020BatchAwareUM,Shao2022Oversubscription,Knap2019UM}. Recent work on Grace--Hopper and AMD MI300A further demonstrates the importance of placement and data movement in tightly coupled CPU--GPU
systems \cite{Fusco2024GH200DataMovement,Schieffer2024GH200SystemMemory, Wahlgren2025MI300AUPM}. GPU microbenchmarking has also exposed architectural behavior across generations, including Hopper and Blackwell
\cite{Volkov2008BenchmarkingGPUs,Wong2010Demystifying,Mei2017GPUmemory, Jia2018Volta,Jia2019Turing,Luo2024Hopper,Jarmusch2025BlackwellMicrobenchmarks, Jarmusch2025BlackwellInDepth}. However, existing Blackwell studies target local GPU microarchitecture, while Grace--Hopper studies characterize the previous generation. Grace--Blackwell memory-management behavior remains insufficiently characterized.

This paper studies NVIDIA Grace--Blackwell UM using placement-controlled microbenchmarks that isolate GPU-local HBM, host/system-memory access, explicit movement, managed-memory placement, access-pattern sensitivity, oversubscription, and CPU--GPU handoff. GB200 is our primary target, with GH200 providing a Grace--Hopper predecessor baseline and GB10/DGX Spark used for benchmark and workflow validation.

Our measurements show that GB200 unified-memory performance is highly state-dependent. CPU-initialized first touch is dominated by runtime page management, whereas prefetched and reused managed memory can recover near-HBM performance. Access pattern further shapes this behavior: streaming accesses amortize memory-management costs effectively, while strided, random, and sparse page-like accesses reduce useful-byte efficiency and expose larger runtime overheads. Under capacity pressure, managed memory remains near resident performance below the GPU-memory boundary but degrades sharply once migration and eviction dominate execution. To explain these effects, we develop a measurement-calibrated analytical model that decomposes execution time into compute, memory-access, and runtime memory-management costs. The model captures when UM behaves like local GPU memory and when explicit placement or locality-aware restructuring remains beneficial.

This paper makes the following contributions:
\begin{itemize}
    \item We design a placement-controlled CUDA microbenchmark suite that
    separates GPU-local HBM, host/system-memory access, explicit movement,
    managed-memory placement, access-pattern sensitivity, oversubscription,
    and CPU--GPU handoff.

    \item To the best of our knowledge, we present one of the first GB200
    unified-memory characterizations, with GH200 and GB10/DGX Spark evaluated
    using a common methodology.

    \item We show that GB200 managed-memory performance depends strongly on
    residency, access history, locality, reuse, and capacity pressure.

    \item We develop a measurement-calibrated analytical model separating
    compute, local and remote memory access, page management, migration,
    eviction, synchronization, and CPU--GPU handoff.
\end{itemize}

Although UM is well established, to the best of our knowledge its behavior on GB200 has not been characterized in a placement-controlled, system-level study. We show that GB200 improves resident and prefetched bandwidth relative to GH200, while cold first touch, CPU-memory access, irregular locality, and oversubscription remain major limits. By isolating and modeling these costs, we identify when managed memory approaches local-HBM performance and when placement, data movement, or capacity pressure dominates. The microbenchmark suite and supporting scripts are withheld during double-blind review and will be open-sourced after the review process.

\section{Background and Motivation}

\begin{figure}[!t]
\centering
\includegraphics[width=\columnwidth]{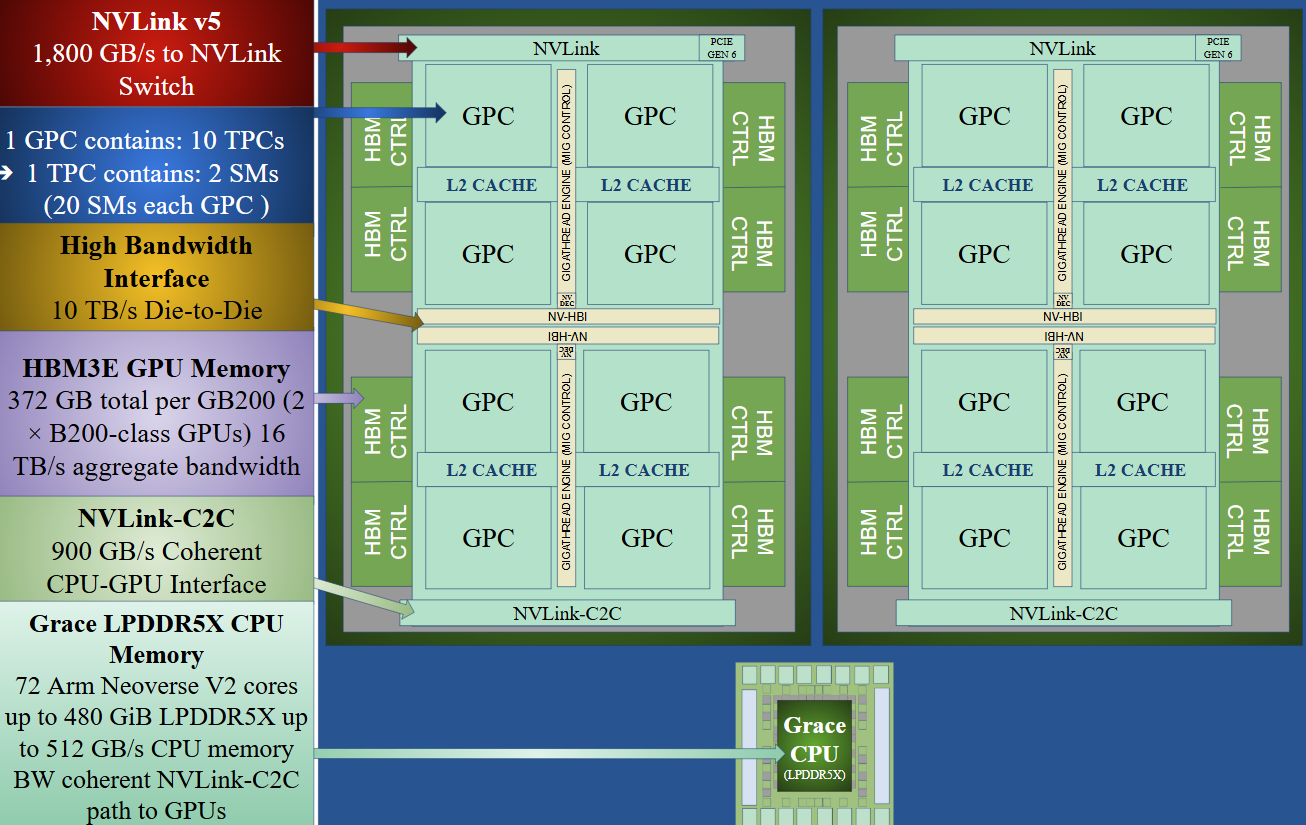}
\caption{GB200 Grace--Blackwell memory organization.}
\label{fig:gb200-architecture}
\end{figure}

\subsection{Grace--Blackwell Memory Paths}

Modern GPU nodes expose multiple memory paths with different capacity, bandwidth, latency, coherence, and placement properties. GPU-local HBM is optimized for high-throughput device execution and provides the fastest memory path for GPU kernels. CPU-attached memory provides larger system capacity and is managed by the host processor, but GPU access to that memory may require explicit transfers, runtime-managed migration, or remote/coherent access through the CPU--GPU interconnect. In traditional PCIe-attached systems, programmers commonly use explicit allocation and copy operations to move data into GPU memory before launching kernels. This approach gives direct placement control, but it also increases programming complexity and makes performance sensitive to transfer timing, redundant copies, and GPU-memory capacity limits \cite{Li2015UnifiedMemory,Harris2013CUDA6UM}.

Grace--Blackwell changes this organization by tightly coupling Grace CPU memory and Blackwell GPU HBM through NVIDIA NVLink-C2C. GB200-class systems combine Grace CPUs and Blackwell GPUs within a coherent CPU--GPU platform, while larger configurations such as GB200 NVL72 connect multiple Grace--Blackwell components through NVLink and NVLink Switch infrastructure \cite{NVIDIA_NVLINK_C2C,NVIDIA_GB200_NVL72,NVIDIA_NVLINK_SYSTEMS}. This organization reduces the programming gap between host and device memory, but it does not remove the performance differences between memory locations. A GPU load served from local HBM, a GPU load satisfied from CPU-attached memory, a managed page migrated into GPU memory, and an oversubscribed page reloaded after eviction can each exercise different hardware paths and runtime mechanisms.

Figure~\ref{fig:gb200-architecture} illustrates the memory paths that motivate our placement-controlled methodology. We therefore treat Grace--Blackwell memory as a set of measurable paths rather than as a uniform memory pool.

\subsection{Addressing, Managed Memory, and Coherent Access}

The term ``UM'' is often used broadly, but several layers contribute to application-visible behavior. We distinguish unified virtual addressing, CUDA managed memory, and the hardware-level coherent CPU--GPU memory system.

Unified virtual addressing provides a common virtual address space in which CPU and GPU pointers can coexist. This simplifies pointer use across host and device code, but it is primarily an addressing mechanism. It does not guarantee that data is physically located near the processor accessing it, that a page has migrated to GPU memory, or that CPU-attached memory and GPU HBM have the same latency or bandwidth.

CUDA UM, or managed memory, is the runtime and driver layer exposed through APIs such as \texttt{cudaMallocManaged}, \texttt{cudaMemPrefetchAsync}, and \texttt{cudaMemAdvise} \cite{NVIDIA_CUDA_UM_2025}. Managed memory allows CPU and GPU code to access the same allocation while the CUDA runtime manages page mapping, placement, migration, access permissions, and eviction. Unlike unified virtual addressing alone, managed memory can introduce runtime actions such as page-fault servicing, prefetching, migration between memory domains, and page replacement. Its performance therefore depends on where pages are resident, whether data has been touched by the CPU or GPU, whether prefetching or advice is used, and whether pages are reused.

The hardware-level coherent memory system defines the physical paths available on the platform. On Grace--Blackwell, this includes Grace CPU memory, Blackwell GPU HBM, NVLink-C2C CPU--GPU connectivity, NVLink GPU--GPU connectivity, and coherence mechanisms that enable tighter CPU--GPU sharing than in conventional PCIe-attached systems \cite{NVIDIA_NVLINK_C2C,NVIDIA_Blackwell_TechnicalBrief,NVIDIA_GB200_NVL72}. These mechanisms can make sharing and remote access more efficient, but they do not make every memory reference equivalent. Local HBM access, coherent CPU-memory access, migrated managed pages, and oversubscribed managed pages remain distinct performance cases.

This distinction also matters for higher-level programming models. In directive-based models such as OpenACC and OpenMP offload, a unified or coherent memory space can reduce the amount of explicit data-management code required, since CPU and GPU code may refer to the same data without manual staging in every case. However, this programmability benefit does not eliminate data movement: the runtime, compiler, or driver may still need to establish mappings, migrate pages, preserve coherence, or serve accesses through a slower memory path.

\subsection{Runtime Policy, Locality, and Capacity Pressure}

CUDA managed memory exposes several runtime behaviors that directly affect performance. When a GPU accesses a managed page that is not resident or not mapped for GPU access, the system may service a page fault, update page-table mappings, migrate data, and resume execution. Programmers can influence this behavior with \texttt{cudaMemPrefetchAsync}, which can move managed pages before kernel execution, and \texttt{cudaMemAdvise}, which provides placement and access hints to the runtime \cite{NVIDIA_CUDA_UM_2025,Sakharnykh2017UM}. Prior work has shown that these mechanisms can improve performance when they reduce demand faults or improve placement, but their effectiveness depends on platform topology, access locality, reuse, and memory pressure \cite{Chien2019CUDAUM,Knap2019UM}.

Access pattern determines whether memory-management costs are amortized efficiently. Streaming access consumes many useful bytes per migrated page or memory transaction, making it favorable for bandwidth-oriented kernels. Strided access reduces useful bytes per cache line, memory transaction, or page. Random and irregular access can reduce coalescing, increase cache and TLB pressure, and make page movement poorly aligned with useful work. Therefore, the same managed allocation can perform differently under streaming, strided, random, and sparse page-like access, even when the total allocation size is unchanged.

Capacity pressure introduces another source of cost. When a managed-memory working set approaches or exceeds GPU-local memory capacity, UM can allow execution to continue without manually restructuring the application around the HBM limit. However, if pages are evicted and later reused, execution can become dominated by repeated migration and reloading rather than useful computation. Prior studies have shown that oversubscription can cause severe slowdowns, especially for irregular workloads and access patterns with poor locality \cite{Shao2022Oversubscription,Allen2021UVM,Kim2020BatchAwareUM}. On Grace--Blackwell, the practical cost of oversubscription depends on the combined behavior of local HBM access, CPU-memory access, page migration, eviction, synchronization, and runtime placement policy.

\subsection{Need for Placement-Controlled Measurement}

Grace--Blackwell motivates placement-controlled characterization because a unified allocation may correspond to local HBM access, CPU-memory access, first-touch migration, prefetched or reused pages, CPU--GPU handoff, or oversubscribed pages undergoing eviction. These cases cannot be explained by one bandwidth number. We therefore isolate local and remote access, explicit movement, managed-memory state, access-pattern sensitivity, handoff, and capacity pressure, and use these measurements to parameterize the analytical model.

\section{Related Work}
\subsection{CUDA UM and Oversubscription}

CUDA UM reduces explicit CPU--GPU data management but introduces runtime costs from page mapping, migration, prefetching, and eviction.
Prior studies show that managed-memory performance depends strongly on prefetching, memory advice, locality, reuse, and whether execution exceeds GPU
memory capacity
\cite{Li2015UnifiedMemory,Harris2013CUDA6UM,Chien2019CUDAUM, Sakharnykh2017UM,Knap2019UM}. Allen and Ge analyze UVM page-fault generation and servicing \cite{Allen2021UVM}, Kim et al. target irregular workloads with batch-aware memory management \cite{Kim2020BatchAwareUM}, and Shao et al. show that oversubscription can degrade sharply once eviction and re-migration dominate execution \cite{Shao2022Oversubscription}.

These studies establish the importance of page faults, locality, prefetching, and capacity pressure, but largely target earlier discrete GPUs or earlier NVLink systems. We instead study these effects on GB200 Grace--Blackwell, where managed memory operates across tightly coupled Grace memory and Blackwell HBM connected by NVLink-C2C.

\subsection{Tightly Coupled CPU--GPU Memory Systems}

Fusco et al. characterize data movement on Grace Hopper using latency, pointer-chasing, copy, ping-pong, internode, and application benchmarks \cite{Fusco2024GH200DataMovement}, establishing GH200 as an important predecessor for tightly coupled Grace--Blackwell systems. Schieffer et al. provide the closest prior work, studying GH200 system-memory behavior including first touch, page migration, page size, oversubscription, and system-, managed-, and GPU-allocated memory \cite{Schieffer2024GH200SystemMemory}.

Our work extends this direction with a direct GB200--GH200 comparison under a common methodology, controlled access-pattern measurements, CPU--GPU handoff and reuse experiments, oversubscription-threshold characterization, and a measurement-calibrated analytical model that separates access and memory-management costs. Cross-vendor work by Wahlgren et al. characterizes Unified Physical Memory on AMD MI300A, including bandwidth, latency, coherence, page faults, TLB behavior, and application implications \cite{Wahlgren2025MI300AUPM}. Although MI300A differs architecturally from Grace--Blackwell, it similarly motivates separating local memory access, CPU--GPU sharing, and runtime memory management rather than treating UM as a uniform resource.

\subsection{GPU Microbenchmarking and Blackwell Characterization}

GPU microbenchmarking has long been used to expose architectural behavior not fully described by vendor documentation, including memory hierarchy, instruction behavior, and cache effects \cite{Volkov2008BenchmarkingGPUs,Wong2010Demystifying,
Mei2017GPUmemory,Jia2018Volta,Jia2019Turing}. More recent studies apply this approach to Hopper and Blackwell \cite{Luo2024Hopper,Luo2025Hopper,Jarmusch2025BlackwellMicrobenchmarks,
Jarmusch2025BlackwellInDepth}, primarily focusing on local GPU microarchitecture and compute behavior. System-level suites such as NVIDIA HPC Benchmarks provide broader end-to-end measurements \cite{NVIDIA_HPC_Benchmarks}, but do not isolate managed-memory state, CPU/system-memory access, handoff, or oversubscription effects.

Our work instead uses placement-controlled microbenchmarks to isolate GPU-local HBM, CPU/system-memory access, managed-memory state, access-pattern sensitivity, oversubscription, and CPU--GPU handoff on
Grace--Blackwell, then uses those measurements to parameterize an analytical model.

\section{Methodology}
This section describes the platforms, benchmark suite, and measurement strategy used to characterize Grace--Blackwell unified-memory behavior. Our methodology uses targeted CUDA microbenchmarks rather than relying solely on end-to-end applications because controlled kernels can isolate memory placement, access pattern, runtime page management, prefetching, memory advice, CPU--GPU handoff, and capacity pressure. 

\subsection{Experimental Platforms}

\begin{table}[t]
\centering
\caption{Benchmark-relevant hardware and software configurations.}
\label{tab:platform-config}

\setlength{\tabcolsep}{1.4pt}
\renewcommand{\arraystretch}{1.08}

\begin{tabularx}{\columnwidth}
{|>{\raggedright\arraybackslash}p{0.205\columnwidth}|
 >{\raggedright\arraybackslash}X|
 >{\raggedright\arraybackslash}X|
 >{\raggedright\arraybackslash}X|}
\hline

\rowcolor{gray!20}
\textbf{Parameter} &
\textbf{GB200} &
\textbf{GH200} &
\textbf{GB10} \\
\hline

\cellcolor{gray!8}\textbf{Driver / CUDA toolkit} &
580.178.04; 13.2.78 &
580.65.06; 13.0.48 &
580.95.05; 13.0.88 \\
\hline

\cellcolor{gray!8}\textbf{CPU} &
Grace; 72 Neoverse-V2 cores &
Grace; 72 Neoverse-V2 cores &
20-core Arm; 10 X925 + 10 A725 \\
\hline

\cellcolor{gray!8}\textbf{CPU caches} &
L1d/i 4.5/4.5 MiB; L2 72 MiB; L3 114 MiB &
L1d/i 4.5/4.5 MiB; L2 72 MiB; L3 114 MiB &
L1d/i 1.3/1.3 MiB; L2 25 MiB; L3 24 MiB \\
\hline

\cellcolor{gray!8}\textbf{Host memory} &
Grace LPDDR5X &
Grace LPDDR5X &
Shared LPDDR5X \\
\hline

\cellcolor{gray!8}\textbf{GPU configuration} &
2$\times$ B200 GPUs; 2 dies/GPU &
1$\times$ Hopper GPU; single die &
1$\times$ integrated Blackwell GPU \\
\hline

\cellcolor{gray!8}\textbf{Benchmark scope} &
Single CUDA device &
Single CUDA device &
Single CUDA device \\
\hline

\cellcolor{gray!8}\textbf{CUDA-visible GPU memory} &
189{,}471 MiB/GPU HBM &
97{,}871 MiB/GPU HBM &
Shared LPDDR5X  \\
\hline

\cellcolor{gray!8}\textbf{CPU--GPU memory organization} &
Grace LPDDR5X + Blackwell HBM via coherent NVLink-C2C &
Grace LPDDR5X + Hopper HBM via coherent NVLink-C2C &
Shared coherent LPDDR5X pool \\
\hline

\end{tabularx}
\end{table}

Table~\ref{tab:platform-config} summarizes the evaluated platforms and measurement scope. A GB200 Superchip couples one Grace CPU with two B200 GPUs, whereas GH200 couples one Grace CPU with one Hopper GPU. The GB200 test host contains two GB200 Superchips, but because our goal is to isolate CPU--GPU memory behavior rather than multi-GPU scaling, CUDA execution is restricted to one complete dual-die Blackwell GPU. This retains the Grace--Blackwell memory paths studied here while avoiding additional multi-GPU communication and placement effects; multi-GPU and NVL72
rack-scale behavior are outside our scope. The handoff experiments additionally pin the host thread to one Grace CPU core. GB200 is the primary target of this study, while GH200 provides the Grace--Hopper predecessor baseline. GB10/DGX Spark is used only for benchmark and workflow validation; because its CPU and GPU share a common
LPDDR5X memory pool, it does not expose the same CPU-memory versus GPU-HBM placement distinction.

\subsection{Benchmark Suite}

The benchmark suite is organized around the memory paths and runtime mechanisms that must be separated to understand Grace--Blackwell UM. Table~\ref{tab:benchmarks} summarizes the benchmark groups. The suite progresses from baseline memory paths to runtime-managed placement, access locality, capacity pressure, and CPU--GPU handoff.

Benchmarks emit CSV-formatted output with common fields for working-set size, trial number, runtime, and bandwidth, together with
benchmark-specific metadata such as memory mode, access pattern,
warm-up count, active element count, useful bytes, and allocation
footprint. The latter fields are especially important for sparse or
irregular accesses, where a large allocation may produce relatively
few useful bytes during a single kernel execution.

\begin{table}[t]
\centering
\caption{Microbenchmark groups used to isolate Grace--Blackwell unified-memory behavior.}
\label{tab:benchmarks}

\scriptsize
\setlength{\tabcolsep}{3pt}
\renewcommand{\arraystretch}{1.08}

\begin{tabularx}{\columnwidth}
{|>{\raggedright\arraybackslash}p{0.31\columnwidth}|
 >{\raggedright\arraybackslash}X|}
\hline

\rowcolor{gray!20}
\textbf{Benchmark group} &
\textbf{What it isolates} \\
\hline

\cellcolor{gray!8}\textbf{HBM bandwidth} &
Local GPU-resident bandwidth. \\
\hline

\cellcolor{gray!8}\textbf{Host/system memory} &
GPU access to CPU-backed memory through a mapped host pointer. \\
\hline

\cellcolor{gray!8}\textbf{Transfer / prefetch} &
Explicit copy bandwidth and managed-memory prefetch bandwidth. \\
\hline

\cellcolor{gray!8}\textbf{Managed memory} &
First touch, prefetch, advice, and reused/resident behavior. \\
\hline

\cellcolor{gray!8}\textbf{Access patterns} &
Streaming, strided, random, and sparse memory-access effects. \\
\hline

\cellcolor{gray!8}\textbf{Oversubscription} &
Capacity pressure, migration, eviction, and reloading behavior. \\
\hline

\cellcolor{gray!8}\textbf{CPU--GPU handoff} &
Handoff cost and the effect of repeated GPU reuse across memory policies. \\
\hline

\end{tabularx}
\end{table}

\subsection{Memory Placement and Runtime Policies}

We evaluate several memory placement and runtime-policy conditions. Explicit GPU-local memory uses \texttt{cudaMalloc} and \texttt{cudaMemcpy} to place data in GPU HBM before kernel execution. Kernel time is measured after placement, so this mode provides the device-resident baseline.

Managed memory uses \texttt{cudaMallocManaged} and CPU initialization before GPU access. In this mode, the measured kernel may include page-fault servicing, page-table updates, migration, and placement overhead. Managed prefetch uses \texttt{cudaMemPrefetchAsync} before the timed kernel and records prefetch time separately when supported. Managed advice uses \texttt{cudaMemAdvise} to evaluate preferred GPU location, accessed-by GPU, and preferred CPU location hints \cite{NVIDIA_CUDA_UM_2025}. Managed reuse performs a prior GPU access before measurement to distinguish cold first-touch behavior from reused GPU-resident behavior. Host/system-memory access is measured using pinned mapped host memory allocated with \texttt{cudaHostAlloc} and accessed on the GPU through a device pointer obtained with \texttt{cudaHostGetDevicePointer}. This path measures GPU access to CPU-backed memory rather than explicit copying or managed migration.

\subsection{Access Patterns, Oversubscription, and Handoff}

Access-pattern benchmarks separate memory-management overhead from locality effects. Streaming access touches contiguous elements and represents the best case for coalesced bandwidth. Strided access touches memory at a fixed interval and reduces useful bytes per cache line, memory transaction, or page. Random access uses a shuffled index array to represent irregular workloads. Page-touch access touches one element at 4 KiB, 64 KiB, and 2 MiB intervals. These intervals provide controlled sparse-access patterns for studying locality at different granularities and are not interpreted
as direct measurements of the runtime's internal page size.

Oversubscription experiments allocate managed-memory working sets as a percentage of CUDA-visible memory. For array-based kernels, the reported footprint includes all arrays used by the benchmark. We evaluate streaming, strided, and random oversubscription to identify where performance changes from bandwidth-oriented execution to behavior dominated by page migration, eviction, and reloading. Random oversubscription results are interpreted carefully because the shuffled index array adds metadata storage and increases total memory pressure.

The CPU--GPU handoff benchmark measures alternating CPU and GPU access to the same data under managed, prefetched managed, host-mapped, and explicit-copy policies. It compares CPU-only, GPU-only, and alternating CPU--GPU execution while varying the number of GPU passes performed after each CPU update. This benchmark quantifies how increasing GPU reuse amortizes CPU--GPU handoff and data-movement costs.

\subsection{Measurement Strategy and Limitations}

All reported configurations use repeated trials. The primary
managed-memory, transfer/prefetch, access-pattern, and oversubscription benchmarks use five warm-up iterations to reduce one-time CUDA runtime and first-launch effects before measurement. Full-suite measurements use 25 trials per GB200 configuration and 30 per GH200 and GB10 configuration. The handoff experiments use five warm-up iterations followed by 20 measured trials per pattern across three repetitions. For each configuration, we report median runtime and median useful bandwidth, with minimum, maximum, and standard deviation where appropriate. Median values are emphasized because they are less sensitive to transient system activity and initialization outliers.

For explicit GPU-local benchmarks, transfer time is reported separately from kernel time. For managed-memory benchmarks, measured kernel time includes page mapping, migration, and placement work that occurs during kernel execution. In prefetch mode, prefetch time is recorded separately and kernel time is measured after prefetch completion. For strided, random, and page-touch benchmarks, bandwidth is reported using useful bytes touched by the kernel, while the full allocation footprint is also recorded.

We use Nsight Systems to validate CUDA API activity, kernel launches, explicit copies, and UM events when available \cite{NVIDIA_NsightSystems_UserGuide}. When page-fault, migration, eviction, or hardware-counter data are unavailable because of tool or permission limits \cite{NVIDIA_NsightCompute_ERR_NVGPUCTRPERM}, we infer behavior from runtime, bandwidth, prefetch timing, access pattern, and oversubscription trends. Random-access benchmarks also require index storage, which can increase memory pressure, particularly under oversubscription. Despite these constraints, the suite still separates the major costs used by the analytical model: local and remote access, explicit movement, managed-memory placement, access locality, CPU--GPU handoff, and capacity pressure.

\section{Characterizing the Grace-Blackwell Memory Hierarchy}
This section establishes the baseline memory paths used to interpret the managed-memory, access-pattern, and oversubscription results in Sections~VI and VII. We compare HBM, host-memory, and explicit-copy paths, then show that one streaming kernel runs near HBM bandwidth on resident, prefetched, or reused data but far slower on first-touch managed pages.

\begin{figure}[t]
\centering
\includegraphics[width=\columnwidth]{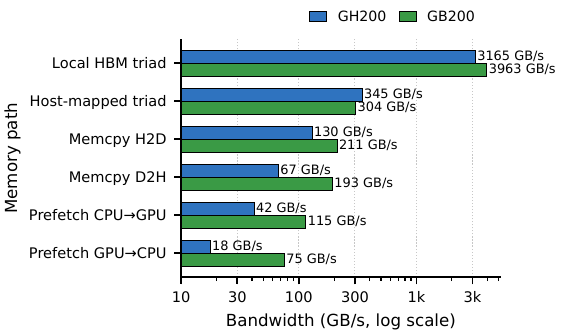}
\caption{Baseline 1~GiB memory-path bandwidths on GH200 and GB200.} 
\label{fig:gh200-gb200-baseline-paths}
\end{figure}

\subsection{Local HBM and Host/System-Memory Access}

Local GPU HBM provides the highest-throughput memory path for GB200 kernels. For a 1~GiB STREAM-style triad, GB200 reaches a median useful bandwidth of 3963~GB/s. The same benchmark on GH200 reaches 3165~GB/s, so the GB200 local-HBM path provides approximately 1.25$\times$ higher measured bandwidth for this kernel. This serves as the baseline for evaluating the remaining memory placement modes.

GPU access to host/system memory is substantially slower than local HBM, but still provides hundreds of GB/s of useful bandwidth. On GB200, the host-mapped 1~GiB triad reaches 304~GB/s, or about 7.7\% of the local-HBM triad bandwidth, compared with 345~GB/s on GH200. Thus, unlike the local-HBM, copy, and prefetch paths, the host-mapped path does not improve on GB200 in this benchmark; its measured bandwidth is about 12\% lower than on GH200. Because this path traverses coherent CPU/system memory rather than local HBM, the result should not be interpreted as a regression in Blackwell HBM performance, and our measurements do not isolate the specific interconnect, coherence, or memory-system factor responsible. It nevertheless reinforces that host/system-memory access is a distinct performance path and should not be treated as equivalent to local HBM for bandwidth-bound kernels.

The copy-only HBM baseline also shows that measured performance depends on the kernel's memory traffic. On GB200, the 1~GiB HBM copy benchmark reaches 2958~GB/s, while the triad reaches 3963~GB/s. We therefore use kernel-specific baselines when normalizing later results instead of assuming a single HBM bandwidth number for all access patterns.

\subsection{Explicit Movement and Managed Prefetch Bandwidth}

Explicit transfer and prefetch measurements provide a movement-bandwidth baseline for interpreting managed-memory behavior. On GB200, 1~GiB explicit host-to-device copies reach a median bandwidth of 211~GB/s, while device-to-host copies reach 193~GB/s. These transfer paths are faster than the corresponding GH200 measurements, where host-to-device and device-to-host copies reach 130~GB/s and 67~GB/s, respectively.

Managed prefetch bandwidth is lower than explicit copy bandwidth but still provides an important mechanism for moving page-management work outside the timed kernel. On GB200, CPU-to-GPU managed prefetch reaches 115~GB/s for a 1~GiB allocation, while GPU-to-CPU prefetch reaches 75~GB/s. On GH200, the corresponding prefetch directions reach 42~GB/s and 18~GB/s. These measurements show that prefetching is not free, but also that GB200 substantially improves the measured page-movement baseline compared with GH200. This distinction is important because later managed-memory results separate the cost of moving pages before a kernel from the cost of servicing faults during kernel execution. Figure~\ref{fig:gh200-gb200-baseline-paths} summarizes these baseline memory paths for the 1~GiB measurements. The results show that GB200 improves local-HBM, explicit-transfer, and managed-prefetch bandwidth relative to GH200, but GPU access to host/system memory remains much slower than local HBM on both platforms.

\subsection{Placement Sensitivity for a Streaming Kernel}

The strongest placement effect appears when comparing the same streaming computation under different residency conditions. On GB200, the explicit GPU-local streaming benchmark at 256~MiB reaches 4582~GB/s. If the allocation is instead created with \texttt{cudaMallocManaged}, initialized on the CPU, and then first accessed by the GPU, bandwidth drops to 14.1~GB/s. This cold first-touch case achieves only 0.3\% of the explicit-HBM baseline because the timed kernel includes runtime page mapping and migration work.

Prefetching or reusing the managed allocation largely eliminates the first-touch penalty. With \texttt{cudaMemPrefetchAsync} before the kernel, the same 256~MiB streaming benchmark reaches 4606~GB/s on GB200, effectively matching the explicit-HBM baseline within measurement variation. When the managed allocation is reused after an initial GPU access, the benchmark reaches 4504~GB/s, or 98.3\% of the explicit-HBM baseline. Thus, the programming abstraction is the same, but the performance depends on whether the managed pages are cold, prefetched, or already resident near the GPU.

\begin{figure}[t]
\centering
\includegraphics[width=\columnwidth]{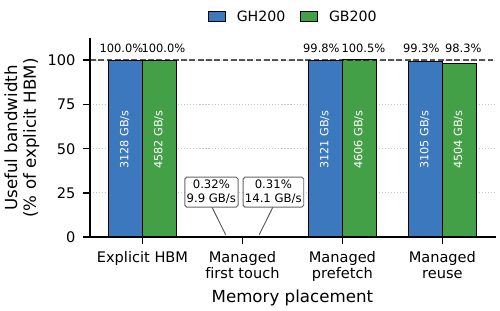}
\caption{Placement sensitivity for the same 256~MiB streaming kernel on GH200 and GB200, normalized to each platform's explicit-HBM baseline.}
\label{fig:gh200-gb200-placement}
\end{figure}

\begin{figure}[t]
\centering
\includegraphics[width=\columnwidth]{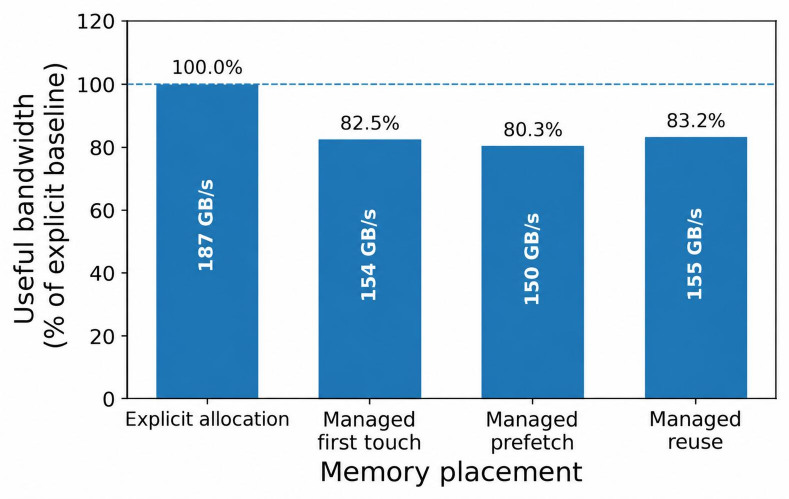}
\caption{GB10/DGX Spark placement-sensitivity validation for a 256~MiB streaming kernel.} 
\label{fig:gb10-placement-validation}
\end{figure}

Figure~3 shows the same qualitative behavior on GH200 and GB200. At 256~MiB, GH200 reaches 3128~GB/s with explicit HBM, 9.9~GB/s under cold first touch, 3121~GB/s with prefetch, and 3105~GB/s with reuse. Thus, prefetch and reuse recover more than 99\% of explicit-HBM bandwidth, while cold first touch retains only 0.32\%. GB200 raises the absolute resident bandwidth but preserves this contrast, showing that residency rather than the managed allocation itself determines streaming performance.

Figure~\ref{fig:gb10-placement-validation} shows the same placement experiment on GB10/DGX Spark as a validation platform. The GB10 explicit allocation baseline reaches 187~GB/s, while managed first touch, managed prefetch, and managed reuse reach 154~GB/s, 150~GB/s, and 155~GB/s, respectively. These cases remain within approximately 80--83\% of the explicit baseline, rather than collapsing as on GH200 and GB200. We therefore use GB10 to validate benchmark behavior, output structure, and workflow reproducibility, while treating GH200 and GB200 as the main architectural comparison for Grace--Hopper and Grace--Blackwell.

Thus, unified addressability does not imply uniform performance: prefetched or reused managed memory can approach local-HBM behavior, while cold first touch and host/system-memory access exercise distinct mechanisms. The next section examines CUDA managed memory in more detail, including how access pattern, memory advice, and reuse affect this placement-sensitive behavior.

\section{CUDA Managed Memory Behavior on Grace-Blackwell}
Section~V showed that GB200 can recover near-HBM bandwidth when managed pages are prefetched or reused, but that cold CPU-initialized managed memory is much slower. Here we separate first touch, prefetching, memory advice, reuse, and access pattern to find when page management dominates.

\subsection{First Touch, Prefetch, Reuse, and Advice}

The default managed-memory case is dominated by first-touch behavior when data is initialized on the CPU and first accessed by the GPU. On GB200, the 256~MiB managed size sweep reaches a median bandwidth of 14.1~GB/s in the cold managed case. At 1~GiB, the same case remains near 14.2~GB/s, indicating that the timed kernel is limited by runtime page mapping and migration rather than by local HBM bandwidth.

Prefetching changes this behavior substantially. With \texttt{cudaMemPrefetchAsync}, the 256~MiB managed sweep reaches 4606~GB/s, while the 1~GiB case reaches 4826~GB/s. Managed reuse shows similar behavior after the initial GPU access: excluding the cold first iteration, reuse reaches 4504~GB/s at 256~MiB and 4751~GB/s at 1~GiB. These results show that the same \texttt{cudaMallocManaged} allocation can behave like a low-bandwidth migration path during first touch or like GPU-resident memory after prefetching or reuse.

Memory advice behaves differently from prefetching. In the 256~MiB advice-mode benchmark, \texttt{cudaMemAdvise} with preferred GPU location reaches 14.4~GB/s, similar to the cold managed case, while preferred CPU location reaches 10.6~GB/s. The accessed-by GPU hint improves bandwidth to 272~GB/s, but remains far below the prefetch and reuse cases, which reach 4809~GB/s and 4746~GB/s in the same benchmark. This indicates that placement hints alone do not necessarily make pages resident before the kernel executes. On GB200, explicit prefetching and prior GPU reuse are the mechanisms that most reliably convert managed memory into near-HBM behavior for streaming access.

\subsection{Access-Pattern Sensitivity}

\begin{figure}[t]
\centering
\includegraphics[width=\columnwidth]{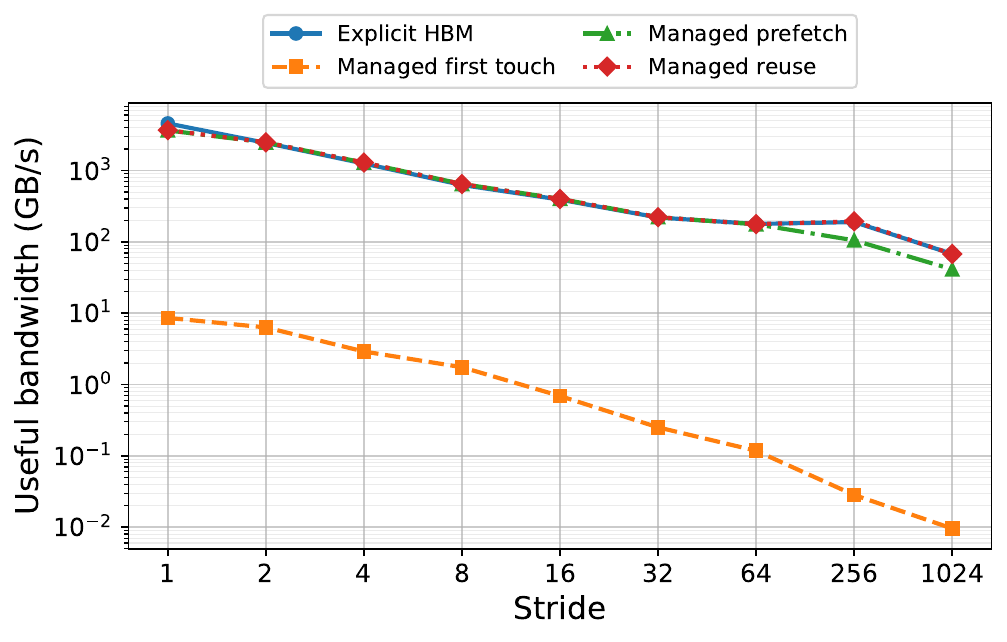}
\caption{GB200 strided-access behavior for a 256~MiB allocation.}
\label{fig:gb200-strided-managed}
\end{figure}

Managed-memory performance also depends strongly on access pattern. Here, stride denotes the number of array elements between consecutive accesses. The allocation footprint remains fixed at 256 MiB, but increasing stride reduces the number of elements actually touched by the kernel. Consequently, each cache line, memory transaction, or managed-memory region may contribute fewer useful bytes to the computation. Because we report useful bandwidth using only the bytes associated with accessed elements, decreasing spatial locality appears directly as lower useful bandwidth.

Figure~5 compares explicit HBM, cold managed memory, managed prefetch, and managed reuse for the GB200 strided-access benchmark. The explicit-HBM baseline falls from 4585 GB/s at stride 1 to 67.7 GB/s at stride 1024. This decline occurs even without managed-memory migration because increasingly sparse accesses make less effective use of each memory transaction and reduce the amount of useful data transferred per unit time. Prefetched and reused managed memory largely follow this explicit baseline: at stride 1, they reach 3659 GB/s and 3685 GB/s, and at stride 64 they reach 177 GB/s and 177 GB/s, respectively.

Cold managed first touch behaves very differently. At stride 1, cold managed memory reaches only 8.6 GB/s, and it falls to 0.01 GB/s by stride 1024. Sparse accesses compound the locality loss because the runtime may still incur page mapping and migration work for memory regions from which only a small fraction of bytes are actually used. As stride increases, less useful work is therefore performed for each unit of memory-management and memory-system traffic, amplifying the first-touch penalty.

Figure~6 shows the corresponding full-stride behavior on GH200. Explicit HBM falls from 3107 GB/s at stride 1 to 78.3 GB/s at stride 1024 as locality decreases. Managed prefetch and reuse generally track the explicit baseline, reaching 3020 and 3047 GB/s at stride 1 and 118 and 114 GB/s at stride 64, respectively. Cold managed first touch again behaves very differently, falling from 8.27 GB/s at stride 1 to 0.008 GB/s at stride 1024. At the largest stride, managed prefetch drops to 45.8 GB/s while explicit HBM and managed reuse remain near 78--82 GB/s. Overall, GH200 and GB200 show the same qualitative access-pattern effect: poorer spatial locality reduces useful-byte efficiency, while cold managed memory additionally incurs page-management and migration overhead.

\begin{figure}[t]
\centering
\includegraphics[width=\columnwidth]{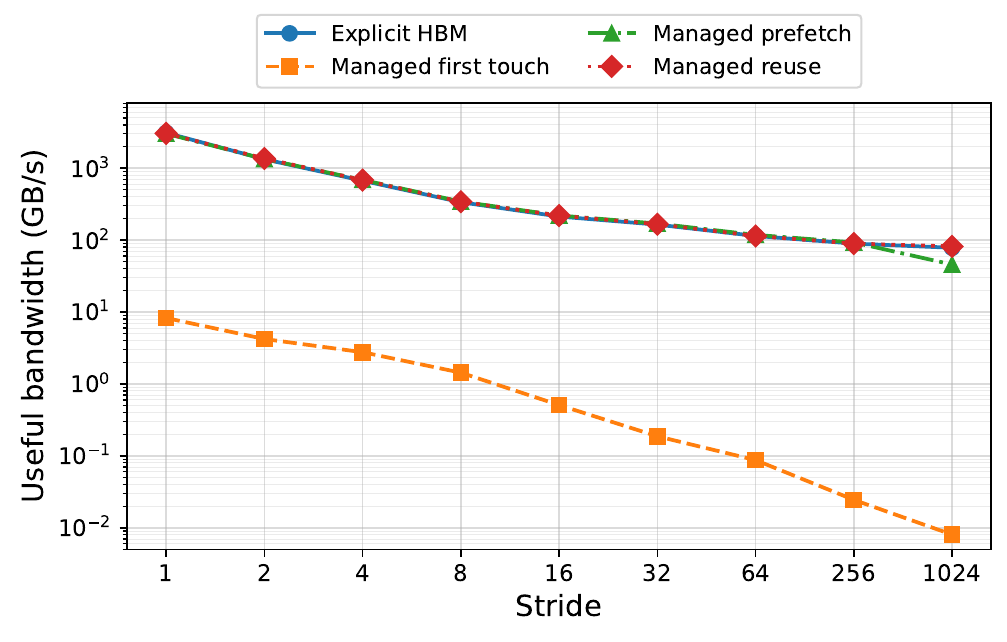}
\caption{GH200 strided-access behavior for a 256 MiB allocation.}
\label{fig:gh200-selected-strided}
\end{figure}

Random access shows the same separation between locality cost and managed-memory placement cost. For the 256~MiB sparse random benchmark, explicit GPU-local memory reaches 144.7~GB/s. Cold managed memory reaches only 0.77~GB/s, while prefetched and reused managed memory recover to 212.8~GB/s and 212.2~GB/s, respectively. The ordering among the resident modes is benchmark-dependent and is not
interpreted as a direct ranking of their underlying memory paths; the important result is that prefetching and reuse remove the dominant first-touch penalty.

\subsection{Sparse Page-Like Access}

The page-touch benchmark further illustrates the importance of useful-byte efficiency. The benchmark touches one element per page-like stride while reporting both useful bytes and allocation footprint. For a 256~MiB managed allocation, a 4~KiB stride reaches 27.0~GB/s of useful bandwidth, while a 64~KiB stride reaches 1.85~GB/s and a 2~MiB stride reaches only 0.043~GB/s. These numbers should not be interpreted as direct measurements of the runtime's internal page size. Instead, they show how sparse access patterns reduce the amount of useful work performed per allocation footprint.

Together, these results show that CUDA managed memory on GB200 is not characterized by a single bandwidth value. Cold CPU-initialized managed memory is dominated by page-management and migration costs. Prefetching and reuse can recover near-resident performance for streaming and regular access. However, strided, random, and sparse page-like patterns reduce useful-byte efficiency and can expose large gaps between the programming abstraction and the underlying memory-management cost.

\section{Capacity Pressure and CPU--GPU Handoff}
Sections~V and VI show that GB200 unified-memory behavior depends on
placement, access history, and locality. This section examines two additional system-level effects: capacity pressure as the working set approaches or exceeds GPU memory, and CPU--GPU handoff when the same data alternates between processors. These measurements expose the capacity- and handoff-related costs represented in the analytical model of Section~VIII; the handoff sweep also provides a directly testable case for quantitative model validation.

\subsection{Capacity Pressure and Oversubscription}

\begin{figure*}[t]
\centering
\includegraphics[width=\textwidth]{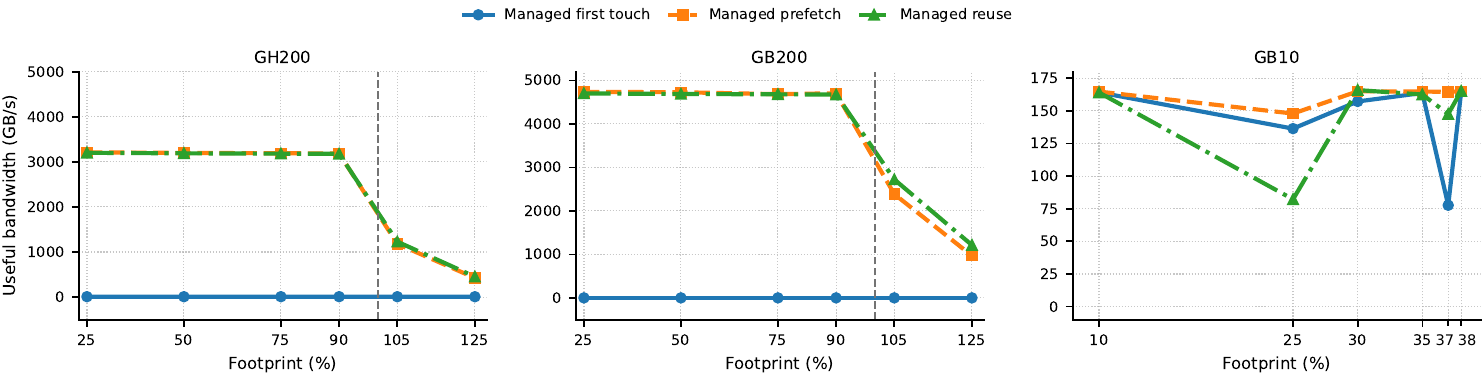}
\caption{Streaming oversubscription behavior across GH200, GB200, and GB10/DGX Spark.}
\label{fig:platform-oversubscription}
\end{figure*}

Managed memory can preserve programmability when the working set approaches or exceeds GPU memory capacity, but performance changes sharply once pages must be evicted and later reloaded. Figure~\ref{fig:platform-oversubscription} compares streaming oversubscription behavior across GH200, GB200, and GB10/DGX Spark. The GH200 and GB200 measurements show the same qualitative threshold behavior: prefetched and reused managed memory remain near resident bandwidth below the capacity boundary, then drop sharply once the managed footprint exceeds CUDA-visible memory.

On GB200, prefetched managed memory sustains 4683--4736~GB/s from 25\% to 90\% of CUDA-visible memory, while reused managed memory sustains 4674--4701~GB/s over the same range. These values are close to the resident streaming bandwidth observed in Sections~V and VI. At 105\%, prefetched managed bandwidth drops to 2385~GB/s and reused managed bandwidth drops to 2722~GB/s. At 125\%, the same modes fall further to 968~GB/s and 1214~GB/s. This drop indicates that execution is no longer limited primarily by streaming HBM bandwidth and is consistent with increasing page migration, eviction, and reloading costs.

GH200 shows the same trend at lower absolute bandwidth. From 25\% to 90\% of CUDA-visible memory, prefetched and reused managed memory remain near 3.2~TB/s. At 105\%, they fall to approximately 1180~GB/s and 1229~GB/s, and at 125\% they fall to 424~GB/s and 453~GB/s. Thus, GB200 increases resident and prefetched streaming bandwidth relative to GH200, but both systems expose the same capacity-pressure transition once the working set exceeds GPU memory capacity.

In contrast, cold managed first-touch bandwidth remains low across the sweep on both GH200 and GB200 because it is already dominated by runtime page-management overhead before oversubscription becomes the main limiting factor. The GB10/DGX Spark data is included as validation evidence: it confirms that the benchmark workflow exercises the same managed-memory policies, but we do not use it to draw final Grace--Blackwell conclusions because GB10 has a different memory organization and much lower absolute bandwidth. The sharp transition is therefore tied to capacity pressure and data
residency rather than simply to the use of managed memory.

\subsection{CPU--GPU Handoff and Reuse}

Many heterogeneous applications alternate between CPU-side updates and GPU-side computation. To measure this behavior, our handoff benchmark compares CPU-only time, GPU-only time, and alternating CPU--GPU execution under managed, prefetched managed, host-mapped, and explicit-copy policies. The key question is how much GPU reuse is needed before moving data into HBM is worth the cost.

\begin{figure}[t]
\centering
\includegraphics[width=\columnwidth]{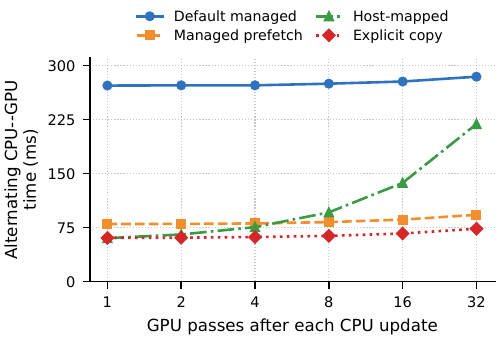}
\caption{GB200 CPU--GPU handoff behavior for a 1~GiB working set as GPU reuse increases after each CPU update.}
\label{fig:gb200-handoff-reuse}
\end{figure}

Figure~\ref{fig:gb200-handoff-reuse} shows the handoff/reuse tradeoff for a 1~GiB working set. With one GPU pass after a CPU update, default managed memory is expensive: alternating execution takes 272.1~ms, compared with a CPU-only plus GPU-only baseline of 54.2~ms. The additional 217.9~ms reflects the cost of demand-driven page management and data movement. Managed prefetch reduces this cost substantially: alternating execution takes 79.9~ms, with 24.6~ms of extra cost over the corresponding baseline. Explicit copy takes 60.5~ms, only 5.4~ms above baseline. Host-mapped memory takes 60.2~ms and adds little handoff cost relative to its matched baseline, but its GPU-only pass is much slower because the GPU accesses CPU-backed memory directly rather than local HBM.

As reuse increases, the policy tradeoff changes. With 32 GPU passes after each CPU update, explicit copy takes 73.3~ms total, managed prefetch takes 92.6~ms, host-mapped memory takes 218.7~ms, and default managed memory takes 284.4~ms. Thus, host-mapped access is competitive when data is used only once, but local placement becomes increasingly valuable as GPU reuse increases. This behavior motivates a simple break-even view: moving data into HBM is worthwhile when the one-time movement cost is repaid by enough faster GPU accesses.

The nearly fixed policy-dependent handoff cost and approximately linear growth with GPU reuse suggest an affine model with a one-time fixed cost and a marginal cost per GPU pass. Section~VIII derives this model from the
measured components and evaluates it against the remaining reuse points.

\section{Measurement-Calibrated Analytical Model}
\subsection{Cost Decomposition}
The measurements above show that Grace--Blackwell memory behavior cannot be represented by a single bandwidth term. We decompose execution time into compute, memory-access, and runtime memory-management costs:
\vspace{-0.8em}

\begin{equation}
T_{\mathrm{total}} =
T_{\mathrm{compute}} + T_{\mathrm{access}} + T_{\mathrm{management}} .
\end{equation}

For bandwidth-dominated execution,

\begin{equation}
T_{\mathrm{access}} \approx
\frac{B_{\mathrm{HBM}}}{BW_{\mathrm{HBM}}} +
\frac{B_{\mathrm{CPU}}}{BW_{\mathrm{CPU}}},
\end{equation}

Here, $B_{\mathrm{HBM}}$ and $B_{\mathrm{CPU}}$ denote useful bytes served from GPU-local HBM and CPU/system memory, respectively, while $BW_{\mathrm{HBM}}$ and $BW_{\mathrm{CPU}}$ are the corresponding bandwidths calibrated from the placement-controlled benchmarks. On GB200, the 1~GiB streaming triad reaches 3963~GB/s from local HBM but only 304~GB/s through host-mapped system memory. Runtime-controlled placement contributes additional costs:
\vspace{-1.0em}

\begin{equation}
T_{\mathrm{management}} =
T_{\mathrm{fault}} + T_{\mathrm{migration}} +
T_{\mathrm{eviction}} + T_{\mathrm{sync}} .
\end{equation}

Here, the four terms represent fault servicing, migration, eviction/reloading,
and synchronization or handoff costs, respectively. Cold CPU-initialized managed memory exposes fault servicing and demand-driven migration, while prefetch and reuse reduce these costs by establishing residency before the measured access. The oversubscription sweep exposes the onset of eviction and reloading once the working set exceeds available GPU memory, and the alternating CPU--GPU benchmark
captures policy-dependent handoff and synchronization costs. For example,
GB200 streaming bandwidth falls from 4606~GB/s with prefetched managed
memory to 14.1~GB/s under cold first touch, while oversubscribed prefetch
falls from more than 4.6~TB/s below the capacity boundary to 968~GB/s at
125\% of CUDA-visible memory. These measurements show that the dominant
term changes with residency and capacity rather than remaining constant
for a managed allocation.

\subsection{Reuse-Amortized Model}

The handoff experiment provides a directly testable instance of this decomposition. For policy $p$, working-set size $B$, and $R$ GPU passes after a CPU update, we model alternating execution as
\vspace{-0.8em}

\begin{equation}
\widehat{T}_p(B,R) = F_p(B) + R\,G_p(B),
\end{equation}

where $F_p(B)$ is the fixed CPU-side and handoff cost and $G_p(B)$ is the marginal cost of one GPU pass. We obtain $G_p(B)$ from the matched GPU-only measurements and calibrate

\vspace{-0.8em}
\begin{equation}
F_p(B) = T_p(B,1)-G_p(B)
\end{equation}

from the measured $R=1$ alternating case. The alternating measurements at $R = 2, 4, 8, 16,$ and $32$ are therefore used for validation rather than model calibration. The two parameters have a direct performance interpretation. $F_p$ captures costs paid once per CPU--GPU exchange, including CPU-side work and
policy-dependent handoff or placement overhead, whereas $G_p$ captures the
incremental cost of each additional GPU pass. Thus, policies that place data
in HBM can pay a larger fixed cost but benefit from a small per-pass cost,
while host-mapped access avoids most up-front movement but continues to pay
for remote access on every GPU pass.

\subsection{Model Validation and Implications}
\begin{table}[t]
\centering
\caption{GB200 reuse-model validation for the 1~GiB handoff benchmark.
Errors are evaluated over $R=2,4,8,16,32$.}
\label{tab:model-validation}
\begin{tabular}{lrrrr}
\hline
Policy & $F_p$ & $G_p$ & Med. APE & Max APE \\
 & (ms) & (ms/pass) & (\%) & (\%) \\
\hline
Explicit copy    & 60.07  & 0.415 & 0.02 & 0.08 \\
Host mapped      & 55.09  & 5.119 & 0.05 & 0.10 \\
Managed          & 271.69 & 0.415 & 0.19 & 0.33 \\
Managed prefetch & 79.37  & 0.489 & 0.93 & 2.61 \\
\hline
\end{tabular}
\end{table}

Table~\ref{tab:model-validation} shows that the model closely reproduces the measured GB200 reuse sweep. For the 1~GiB case, explicit copy, host-mapped access, and default managed memory remain within 0.4\% maximum absolute percentage error, while managed prefetch remains within 2.7\%. Across the 16~MiB, 64~MiB, 256~MiB, and 1~GiB working sets, the aggregate median absolute percentage error over the validation reuse counts is approximately 0.40\%. The model also identifies the measured fastest policy for all 20 validation size--reuse combinations. We exclude the smaller 64~KiB and 2~MiB cases from this policy-selection result because fixed launch and timing overheads become comparable to memory access.

The calibrated parameters also predict the observed crossover between host-mapped and explicit placement. Equating the two policy models gives

\begin{equation}
R^{*} =
\frac{F_{\mathrm{explicit}}-F_{\mathrm{host}}}
     {G_{\mathrm{host}}-G_{\mathrm{explicit}}}
\approx 1.06 .
\end{equation}

At 1~GiB, host-mapped access therefore has an advantage for approximately one GPU pass, but its larger marginal cost makes explicit placement preferable with additional reuse. The measurements confirm this prediction: host-mapped execution is slightly faster at $R=1$ (60.2 versus 60.5~ms), whereas explicit copy is faster at $R=2$ (60.9 versus 65.3~ms), with the gap increasing thereafter.

The decomposition and reuse model serve complementary purposes. The former connects placement, migration, eviction, locality, and synchronization to behaviors observed throughout the benchmark suite, while the latter quantitatively predicts the CPU--GPU placement tradeoff. In practice, our results suggest three guidelines: prefetch or establish GPU residency before bandwidth-intensive managed-memory kernels, avoid sustained oversubscription when the working set exceeds available HBM, and prefer explicit or prefetched placement when data will be reused across GPU passes. Host-mapped access is most attractive when GPU reuse is low and avoiding an up-front movement cost is more important than per-pass bandwidth. Together, these results show why unified addressability alone cannot predict performance: the dominant cost changes with placement, locality, capacity pressure, and reuse.

\section{Conclusion}
We characterized NVIDIA Grace--Blackwell UM on GB200 with placement-controlled microbenchmarks, using GH200 as a Grace--Hopper baseline and GB10/DGX Spark for validation. Relative to GH200, GB200 raises resident and prefetched bandwidth, but cold first touch, CPU-memory access, irregular locality, and oversubscription remain major performance limits. On a 1~GiB STREAM triad, local HBM reaches 3963~GB/s, while host-mapped system memory reaches 304~GB/s, 7.7\% of HBM. On a 256~MiB streaming kernel, cold CPU-initialized managed memory reaches only 14.1~GB/s, while prefetched and reused managed memory reach 4606 and 4504~GB/s. Strided, random, and sparse accesses use fewer of the bytes they move and expose larger managed-memory overheads. Prefetched and reused managed memory stay above 4.6~TB/s up to 90\% of CUDA-visible memory, but at 125\% they fall to 968 and 1214~GB/s.

Our analytical model separates compute, local and remote memory access, page management, migration, eviction, and synchronization costs. It also quantitatively captures the reuse-dependent CPU--GPU placement tradeoff: across the larger handoff working sets, the model achieves approximately 0.40\% median absolute percentage error and identifies the measured fastest policy for all 20 validation size--reuse combinations. These results show when prefetching or explicit placement can amortize movement costs and when remote or oversubscribed access becomes the dominant limitation.

\bibliographystyle{IEEEtran}
\bibliography{references}

@manual{NVIDIA_CUDA_UM_2025,
  title        = {{CUDA C++ Programming Guide: Unified Memory}},
  author       = {{NVIDIA Corporation}},
  year         = {2025},
  organization = {NVIDIA Corporation},
  url          = {https://docs.nvidia.com/cuda/cuda-programming-guide/04-special-topics/unified-memory.html},
  note         = {Section 4.1, accessed 2026-02-23}
}

@misc{NVIDIA_GraceHopper_Architecture,
  author       = {{NVIDIA Corporation}},
  title        = {{NVIDIA Grace Hopper Superchip Architecture In-Depth}},
  year         = {2022},
  month        = nov,
  day          = {10},
  organization = {NVIDIA Developer Blog},
  url          = {https://developer.nvidia.com/blog/nvidia-grace-hopper-superchip-architecture-in-depth/},
  urldate      = {2026-05-28}
}

@misc{NVIDIA_NVLINK_C2C,
  author       = {{NVIDIA Corporation}},
  title        = {{NVIDIA NVLink-C2C: Chip Interconnect Technology}},
  year         = {2025},
  organization = {NVIDIA Corporation},
  url          = {https://www.nvidia.com/en-us/data-center/nvlink-c2c/},
  urldate      = {2026-06-23}
}

@misc{NVIDIA_NVLINK_SYSTEMS,
  author       = {{NVIDIA Corporation}},
  title        = {{NVIDIA Multi-Node NVLink Systems}},
  year         = {2025},
  organization = {NVIDIA Corporation},
  url          = {https://docs.nvidia.com/multi-node-nvlink-systems/index.html},
  urldate      = {2026-06-23}
}

@misc{NVIDIA_GB200_NVL72,
  author       = {{NVIDIA Corporation}},
  title        = {{NVIDIA GB200 NVL72}},
  year         = {2025},
  organization = {NVIDIA Corporation},
  url          = {https://www.nvidia.com/en-us/data-center/gb200-nvl72/},
  urldate      = {2026-05-28}
}

@misc{NVIDIA_Blackwell_TechnicalBrief,
  author       = {{NVIDIA Corporation}},
  title        = {{NVIDIA Blackwell Architecture Technical Overview}},
  year         = {2024},
  organization = {NVIDIA Corporation},
  url          = {https://resources.nvidia.com/en-us-blackwell-architecture},
  urldate      = {2026-06-16}
}

@misc{NVIDIA_HPC_Benchmarks,
  author       = {{NVIDIA Corporation}},
  title        = {{NVIDIA HPC Benchmarks}},
  year         = {2026},
  organization = {NVIDIA Corporation},
  url          = {https://docs.nvidia.com/nvidia-hpc-benchmarks/},
  urldate      = {2026-06-23}
}

@misc{Fusco2024GH200DataMovement,
  title         = {{Understanding Data Movement in Tightly Coupled Heterogeneous Systems: A Case Study with the Grace Hopper Superchip}},
  author        = {Fusco, Luigi and Khalilov, Mikhail and Chrapek, Marcin and Chukkapalli, Giridhar and Schulthess, Thomas and Hoefler, Torsten},
  year          = {2024},
  eprint        = {2408.11556},
  archivePrefix = {arXiv},
  primaryClass  = {cs.DC},
  url           = {https://arxiv.org/abs/2408.11556}
}

@inproceedings{Schieffer2024GH200SystemMemory,
  author    = {Schieffer, Gabin and Wahlgren, Jacob and Ren, Jie and Faj, Jennifer and Peng, Ivy B.},
  title     = {{Harnessing Integrated CPU-GPU System Memory for HPC: A First Look into Grace Hopper}},
  booktitle = {Proceedings of the 53rd International Conference on Parallel Processing},
  series    = {ICPP '24},
  year      = {2024},
  publisher = {Association for Computing Machinery},
  address   = {New York, NY, USA},
  doi       = {10.1145/3673038.3673110},
  eprint    = {2407.07850},
  archivePrefix = {arXiv},
  url       = {https://arxiv.org/abs/2407.07850}
}

@inproceedings{Li2015UnifiedMemory,
  author    = {Li, Wenqiang and Jin, Guanghao and Cui, Xuewen and See, Simon},
  title     = {{An Evaluation of Unified Memory Technology on NVIDIA GPUs}},
  booktitle = {2015 15th IEEE/ACM International Symposium on Cluster, Cloud and Grid Computing},
  year      = {2015},
  pages     = {1092--1098},
  doi       = {10.1109/CCGrid.2015.105}
}

@misc{Harris2013CUDA6UM,
  author       = {Harris, Mark},
  title        = {{Unified Memory in CUDA 6}},
  year         = {2013},
  month        = nov,
  day          = {18},
  organization = {NVIDIA Developer Blog},
  url          = {https://developer.nvidia.com/blog/unified-memory-in-cuda-6/},
  urldate      = {2026-03-13}
}

@inproceedings{Chien2019CUDAUM,
  author    = {Chien, Steven W. D. and Peng, Ivy B. and Markidis, Stefano},
  title     = {{Performance Evaluation of Advanced Features in CUDA Unified Memory}},
  booktitle = {2019 IEEE/ACM Workshop on Memory Centric High Performance Computing},
  series    = {MCHPC '19},
  year      = {2019},
  pages     = {50--57},
  doi       = {10.1109/MCHPC49590.2019.00014},
  eprint    = {1910.09598},
  archivePrefix = {arXiv},
  url       = {https://arxiv.org/abs/1910.09598}
}

@misc{Sakharnykh2017UM,
  author       = {Sakharnykh, Nikolay},
  title        = {{Maximizing Unified Memory Performance in CUDA}},
  year         = {2017},
  month        = nov,
  day          = {19},
  organization = {NVIDIA Developer Blog},
  url          = {https://developer.nvidia.com/blog/maximizing-unified-memory-performance-cuda/},
  urldate      = {2026-03-13}
}

@inproceedings{Shao2022Oversubscription,
  author    = {Shao, Chuanming and Guo, Jinyang and Wang, Pengyu and Wang, Jing and Li, Chao and Guo, Minyi},
  title     = {{Oversubscribing GPU Unified Virtual Memory: Implications and Suggestions}},
  booktitle = {Proceedings of the 2022 ACM/SPEC International Conference on Performance Engineering},
  series    = {ICPE '22},
  year      = {2022},
  pages     = {67--75},
  publisher = {Association for Computing Machinery},
  address   = {New York, NY, USA},
  doi       = {10.1145/3489525.3511691},
  url       = {https://doi.org/10.1145/3489525.3511691}
}

@inproceedings{Allen2021UVM,
  author    = {Allen, Tyler and Ge, Rong},
  title     = {{In-Depth Analyses of Unified Virtual Memory System for GPU Accelerated Computing}},
  booktitle = {Proceedings of the International Conference for High Performance Computing, Networking, Storage and Analysis},
  series    = {SC '21},
  year      = {2021},
  publisher = {Association for Computing Machinery},
  address   = {New York, NY, USA},
  doi       = {10.1145/3458817.3480855},
  url       = {https://doi.org/10.1145/3458817.3480855}
}

@inproceedings{Kim2020BatchAwareUM,
  author    = {Kim, Hyojong and Sim, Jaewoong and Gera, Prasun and Hadidi, Ramyad and Kim, Hyesoon},
  title     = {{Batch-Aware Unified Memory Management in GPUs for Irregular Workloads}},
  booktitle = {Proceedings of the Twenty-Fifth International Conference on Architectural Support for Programming Languages and Operating Systems},
  series    = {ASPLOS '20},
  year      = {2020},
  pages     = {1357--1370},
  publisher = {Association for Computing Machinery},
  address   = {New York, NY, USA},
  doi       = {10.1145/3373376.3378529},
  url       = {https://doi.org/10.1145/3373376.3378529}
}

@article{Knap2019UM,
  author  = {Knap, Maciej and Czarnul, Pawe{\l}},
  title   = {{Performance Evaluation of Unified Memory with Prefetching and Oversubscription for Selected Parallel CUDA Applications on NVIDIA Pascal and Volta GPUs}},
  journal = {The Journal of Supercomputing},
  year    = {2019},
  volume  = {75},
  pages   = {7625--7645},
  doi     = {10.1007/s11227-019-02966-8},
  url     = {https://doi.org/10.1007/s11227-019-02966-8}
}

@misc{Jarmusch2025BlackwellInDepth,
  title         = {{Microbenchmarking NVIDIA's Blackwell Architecture: An In-depth Architectural Analysis}},
  author        = {Jarmusch, Aaron and Chandrasekaran, Sunita},
  year          = {2025},
  eprint        = {2512.02189},
  archivePrefix = {arXiv},
  primaryClass  = {cs.AR},
  url           = {https://arxiv.org/abs/2512.02189}
}

@misc{Jarmusch2025BlackwellMicrobenchmarks,
  title         = {{Dissecting the NVIDIA Blackwell Architecture with Microbenchmarks}},
  author        = {Jarmusch, Aaron and Graddon, Nathan and Chandrasekaran, Sunita},
  year          = {2025},
  eprint        = {2507.10789},
  archivePrefix = {arXiv},
  primaryClass  = {cs.DC},
  url           = {https://arxiv.org/abs/2507.10789}
}

@misc{Wahlgren2025MI300AUPM,
  title         = {{Dissecting CPU-GPU Unified Physical Memory on AMD MI300A APUs}},
  author        = {Wahlgren, Jacob and Schieffer, Gabin and Shi, Ruimin and Le{\'o}n, Edgar A. and Pearce, Roger and Gokhale, Maya and Peng, Ivy B.},
  year          = {2025},
  month         = aug,
  day           = {18},
  eprint        = {2508.12743},
  archivePrefix = {arXiv},
  primaryClass  = {cs.DC},
  url           = {https://arxiv.org/abs/2508.12743}
}

@misc{Luo2024Hopper,
  title         = {{Benchmarking and Dissecting the NVIDIA Hopper GPU Architecture}},
  author        = {Luo, Weile and Fan, Ruibo and Li, Zeyu and Du, Dayou and Wang, Qiang and Chu, Xiaowen},
  year          = {2024},
  eprint        = {2402.13499},
  archivePrefix = {arXiv},
  primaryClass  = {cs.AR},
  url           = {https://arxiv.org/abs/2402.13499}
}

@article{Luo2025Hopper,
  title         = {{Dissecting the NVIDIA Hopper Architecture Through Microbenchmarking and Multiple Level Analysis}},
  author        = {Luo, Weile and Fan, Ruibo and Li, Zeyu and Du, Dayou and Liu, Hao and Wang, Qiang and Chu, Xiaowen},
  journal       = {arXiv preprint arXiv:2501.12084},
  year          = {2025},
  url           = {https://arxiv.org/abs/2501.12084}
}

@article{Mei2017GPUmemory,
  author  = {Mei, Xinxin and Chu, Xiaowen},
  title   = {{Dissecting GPU Memory Hierarchy Through Microbenchmarking}},
  journal = {IEEE Transactions on Parallel and Distributed Systems},
  year    = {2017},
  volume  = {28},
  number  = {1},
  pages   = {72--86},
  doi     = {10.1109/TPDS.2016.2549523}
}

@inproceedings{Wong2010Demystifying,
  author    = {Wong, Henry and Papadopoulou, Mary-Maria and Sadooghi-Alvandi, Maryam and Moshovos, Andreas},
  title     = {{Demystifying GPU Microarchitecture Through Microbenchmarking}},
  booktitle = {2010 IEEE International Symposium on Performance Analysis of Systems and Software},
  year      = {2010},
  pages     = {235--246},
  doi       = {10.1109/ISPASS.2010.5452013}
}

@inproceedings{Volkov2008BenchmarkingGPUs,
  author    = {Volkov, Vasily and Demmel, James W.},
  title     = {{Benchmarking GPUs to Tune Dense Linear Algebra}},
  booktitle = {Proceedings of the 2008 ACM/IEEE Conference on Supercomputing},
  series    = {SC '08},
  year      = {2008},
  publisher = {IEEE Press},
  articleno = {31},
  pages     = {1--11}
}

@misc{Jia2018Volta,
  title         = {{Dissecting the NVIDIA Volta GPU Architecture via Microbenchmarking}},
  author        = {Jia, Zhe and Maggioni, Marco and Staiger, Benjamin and Scarpazza, Daniele P.},
  year          = {2018},
  eprint        = {1804.06826},
  archivePrefix = {arXiv},
  primaryClass  = {cs.DC},
  url           = {https://arxiv.org/abs/1804.06826}
}

@misc{Jia2019Turing,
  title         = {{Dissecting the NVIDIA Turing T4 GPU via Microbenchmarking}},
  author        = {Jia, Zhe and Maggioni, Marco and Smith, Jeffrey and Scarpazza, Daniele Paolo},
  year          = {2019},
  eprint        = {1903.07486},
  archivePrefix = {arXiv},
  primaryClass  = {cs.DC},
  url           = {https://arxiv.org/abs/1903.07486}
}

@misc{NVIDIA_NsightSystems_UserGuide,
  author       = {{NVIDIA Corporation}},
  title        = {{NVIDIA Nsight Systems User Guide}},
  year         = {2026},
  organization = {NVIDIA Corporation},
  url          = {https://docs.nvidia.com/nsight-systems/UserGuide/index.html},
  urldate      = {2026-06-16}
}

@misc{NVIDIA_NsightCompute_ERR_NVGPUCTRPERM,
  author       = {{NVIDIA Corporation}},
  title        = {{{ERR\_NVGPUCTRPERM}: Nsight Compute Permission Issue with Performance Counters}},
  year         = {2026},
  organization = {NVIDIA Developer},
  url          = {https://developer.nvidia.com/nvidia-development-tools-solutions-err-nvgpuctrperm-nsightcompute},
  urldate      = {2026-06-16}
}

\end{document}